\documentclass[a4paper,twocolumn,prb]{revtex4}
\usepackage{amsmath}
\usepackage{graphicx}
\usepackage{color}
\usepackage{subfigure}
\usepackage{hyperref}
\begin{document}
\title{Continuous Move From BTW to Manna Model}
\author {M. N. Najafi }
\affiliation{Physics department, Sharif University of Technology, P.O. Box 11155-9161, Tehran, Iran}
\author{Saman Moghimi-Araghi}

\affiliation{Physics department, Sharif University of Technology, P.O. Box 11155-9161, Tehran, Iran}

\begin{abstract}
        In the present paper we consider the BTW model perturbed by random-direction anisotropy with strength factor $\epsilon$ ranging from 0 to 1 corresponding to BTW and Manna model receptively and investigate the properties of the statistical observables for various rates of anisotropy. By increasing the $\epsilon$, we observe a cross-over taking place between these models. For small length scales, the curves show properties similar to the BTW model whereas in the Infra red limit the corresponding $\kappa$ is nearly the same as the Manna model. The observations confirm that this perturbation is relevant for the BTW fixed point and the infra red limit of the perturbed model is described by the Manna model. We also propose a differential equation whose solution properly fits with the the Green's function obtained by the simulation. This can help us to obtain the action of the perturbed model.
\end{abstract}
\maketitle
\section{Introduction}

     In recent years, the concept of self-organized criticality (SOC) proposed by Bak et. al.\cite{BTW} has attracted a lot of attention as a possible general framework for explanation of the occurrence of robust power laws in nature as it does not require fine tuning of any parameter to set criticality. Sandpile models was the first example of these systems. The abelian structure of the sandpile model was first discovered by D. Dhar and named as Abelian Sandpile Model (ASM) \cite{Dhar2}. Numerous works have been done on this model. The connection of this model with spanning trees\cite{MajDhar2}, ghost models\cite{MahRuel}, q-state Potts model, Loop Erased Random Walk (LERW)\cite{Majumdar} is known. The different height and cluster probabilities of this model\cite{MajDhar} and its various geometrical exponents are also calculated numerically and analytically. For a review see\cite{Dhar3}. Several variations of the ASM have been studied in the past with a view to understand the parameters that determine the different universality classes of self-organized critical behavior. These include models in which particle transfer is directed, or models in which the toppling condition or the number of sand grains transferred depends on the local slope rather than local height\cite{Manna1,KadNagZho,Manna2,Zhang,DharRam}. In this respect, it has been realized that stochasticity in toppling rules can lead to different critical behavior than models with deterministic toppling rules. One of the most interesting BTW-variations is the Manna model\cite{Manna1} in which after a toppling the grains redistribute  randomly in a preferred direction which is randomly chosen (without dissipation). This model corresponds to a two state model in which there is a hard core interaction between two particles in the same site that prevents the site to be doubly occupied. The interesting question then would be what is the universality classes which these models belong to. In fact, the precise identification of universality classes in sandpile models is an unresolved issue. It is generally assumed that the avalanche size and duration distributions follow simple power laws in the infinite-size limit, and the departures from such power laws reflect finite-size effects. Such effects complicate the estimation of critical exponents, since the estimates are sensitive to the choice of fitting interval. In the main Manna's paper \cite{Manna1} it was claimed that the Manna model lies in the same universality class as the BTW model. Real-space renormalization group calculations \cite{PVZ} suggest that different sandpile models, such as the BTW and the Manna models, all belong to the same universality class. This result is also confirmed by a proposed field theory approach\cite{DickVespZap} that states that all sand pile models are described by the same effective field theory at the coarse grained level. Using the new exponents, introduced by K. Christensen et. al. for the sand pile models\cite{ChrisOlami}, A. Ben-Hur et. al. showed that these exponents for Manna model are different from BTW counterparts\cite{BenHur}. They claimed that the Manna model is in the universality class of random neighbor models which is distinct from the BTW universality class. Some more exact numerical results also confirmed this hypothesis\cite{Lubeck}. Based on some numerical analysis (bias removing), A. Chessa et. al. argued that the results of Ben-Hur et. al. are same for the two models which imply that they may belong to the same universality class\cite{Chessa}. They used the finite size scaling (FSS) arguments in calculating the exponents of size, time and area distributions of avalanches of BTW and Manna models which is believed that are not single fractals and do not fulfil the FSS\cite{KitLubGrasPri}. Many papers argued this result and reported some exact results about the scaling behavior of this model\cite{Lubeck,DickCam,HuPruChe}.  
     
In spite of very much works done on these variations, very low attention is paid on the question how these models are linked and transform perturbatively to one another. Knowing the structure of this transition, one can obtain some valuable informations of these models. To this end, one can add perturbatively stochasticity to the BTW model. It can be done by adding anisotropy with random preferable direction with arbitrary strength to the toppling rule of BTW model and see how observables evolve from BTW to the Manna model. A study on patterned and disordered continuous ASM has been done in \cite{AzimiMogh} in which it has been shown that the quenched disorder lead to an irrelevant perturbation in the conformal field theory corresponding to the BTW. The plan of the present paper is to numerically see how things change when the disorder is not quenched. For this, we consider the geometrical objects (interfaces) of the perturbed model and some other statistical functions. The properties of the interfaces are directly related to the correlation length of the system. In the discrete set up, in addition to the correlation length, the system has one more characteristic length scale, i. e. the lattice constant. In the scales much smaller than the system correlation length, and larger than the lattice constant, one expects that the statistical features of the curves are properly fitted to the unperturbed one. There is an idea to describe the geometrical interfaces of 2D critical statistical models via growth processes named as SLE\cite{Schramm}. It is a powerful tool to study the macroscopic interfaces of two dimensional systems instead of local fields as is common in ordinary CFT. The result of this idea is a complete classification of probability measures on random curves in (simply connected) domains of the complex plane satisfying two axioms: conformal invariance and the domain Markov property\cite{BauBerKal}. From the correspondence of ASM with the ghost model\cite{MahRuel}, one expects that this model is a c=-2 conformal field theory and with the knowledge of the connection between conformal field theory and SLE\cite{BauBer} one finds that the ASM is related to SLE with $\kappa=2$. This test is done in \cite{Saberi} in which numerically is shown that the boundary of avalanches of ASM is SLE(2). To generate curves running from origin to the infinity from loops, one has to cut the loops horizontally and then send the end point of the curve to the infinity. In this paper we consider perturbed ASM on a cubic lattice and numerically analyze the properties of interfaces of this model. In sections II and III we briefly introduce ASM and SLE respectively. Sections IV and VI also contain some numerical results from loop statistics and SLE on these interfaces respectively.

\section{Introduction to the Abelian Sandpile Model}\label{ASM}
Consider the ASM on a two dimensional square lattice $L\times{L}$. For each site we consider the height variable $h_{i}$ taking its values from the set $\lbrace{1, 2, 3, 4}\rbrace$ which shows the number of grains in this site. So each configuration of the sandpile is given by the set $\lbrace{h_{i}}\rbrace$. The dynamics of this model is as follows; in each step, a grain is added to a random site $i$ i.e. $h_{i}\rightarrow{h_{i}+1}$, then if the resulting height becomes more than 4, the site toppels and loses 4 sands, each of which is transferred to one of four neighbours of the original site. As a result, the neighboring sites may become unstable and topple and a chain of topplings may happen in the system. In the boundary sites, the toppeling causes to one or two sands to leave the system. This process continue until the system reaches to a stable configuration. Now another random site is selected and the sand is released on this site and the process continue. The movement on the space of stable configuration lead the system to fall in a subset of sets of configurations after a finite steps, named as the “recurrent states”. It has been shown that the total number of recurrent states is det$\Delta$ where $\Delta$ is the discrete Laplacian. For details see \cite{Dhar3}. This model can be generalized to other lattice geometries and to off critical set up. For a lattice with $d$ neighboring sites, the toppling occurs when $h_{i}>d$, then the original site will lose $d$ grains and the height of each of its neighbors will increase by 1. It has been shown
that the action corresponding to this model is:
\begin{equation}
S=\int{d^{2}z}(\partial\theta\bar{\partial}\bar{\theta})
\label{ghost}
\end{equation}
where $\theta$ and $\bar{\theta}$ are complex Grassmann variables.
\newline \
\newline \textbf{Waves}; As is mentioned above, the topplings in the ASM can be done in any order. One very useful way to relax is by a succession of waves of topplings. Let the site where the grain is added be $O$. If after addition, $O$ is still stable, the relaxation process is over. If it is unstable, we relax it as follows: topple $O$ once, and then allow the avalanche to proceed by relaxing any unstable sites, without however toppling $O$ again. This constitutes the first wave of toppling. If at the end, site $O$ is still unstable, we allow it to topple once more, and let the other sites relax, until all sites other than $O$ are stable. This is the second wave of toppling. Repeat as needed.
Eventually, site $O$ is no longer unstable at the end of a wave, and the relaxation process stops. It is easy to see
that in any wave, the set of toppled sites forms a connected cluster with no voids (untoppled sites fully
surrounded by toppled sites), and no site topples more than once in one wave. (This would not be true if the
graph had greedy sites). 
\newline \
\newline \textbf{Random Anisotropic ASM}; One can add random anisotropy to ASM in the following sense; Choose random number $r=\pm 1$. When the height of a site becomes more than $4n$ where $n$ is some integer, then $4n$ grains is transferred to the neighboring sites; $n-r$ grains transfer to the 'up' site, $n-r$ grains to the 'down' site, $n+r$ grains to the 'left' site and $n+r$ grains to the 'right' site. In this respect we define $\epsilon=\frac{1}{n}$. $\epsilon=1$ and $\epsilon=0$ correspond to the Manna and BTW models respectively. When $\epsilon=0$ or $\epsilon=1$ the observables have robust power low behaviors up to a characteristic length (named as correlation length $\xi$) above which the correlation functions falls off rapidly. In a critical model, this length is of order of the lattice size and when the system size goes to infinity, it diverges. The correlation length can be defined as the loop linear size ($r_{cut}$) above which the logarithm of the distribution function of gyration radius falls off more rapidly than linear. When $0<\epsilon<1$ it is seen that there is another characteristic length $\xi_{2}$ in which the behavior of the statistical functions smoothly changes.

\section{NUMERICAL RESULTS; STATISTICS OF PERTURBED ASM}\label{stat}

\begin{figure}
\centerline{\includegraphics[scale=.55]{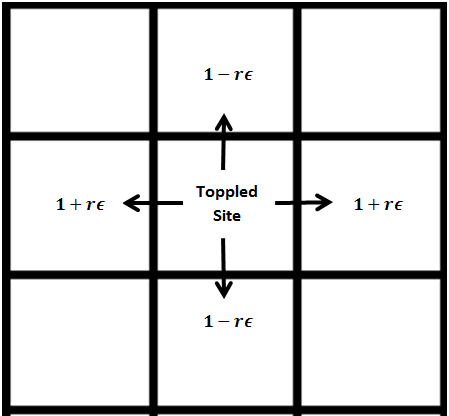}}
\caption{The schematic view of square lattice. The toppling rule is schematically shown (r is the random number taking
values +1 and -1).}
\label{sample}
\end{figure}

\begin{figure}
\centerline{\includegraphics[scale=.35]{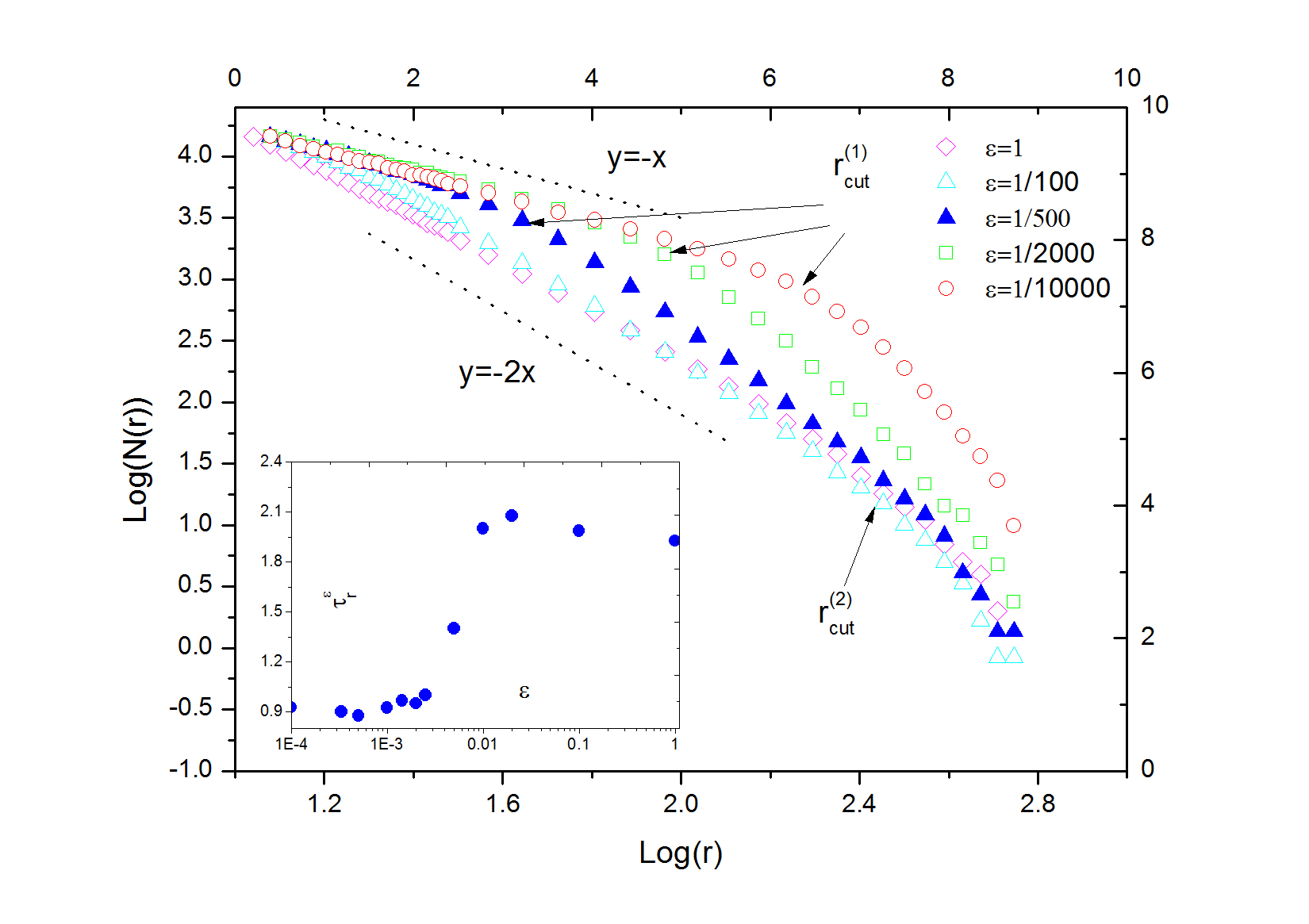}}
\caption{The behaviour of log(N(r)) versus gyration radius Log(r), for various dissipations for waves. The inner graph shows scaling of their slopes in terms of mass.}
\label{gyration}
\end{figure}
\begin{figure}
\centerline{\includegraphics[scale=.35]{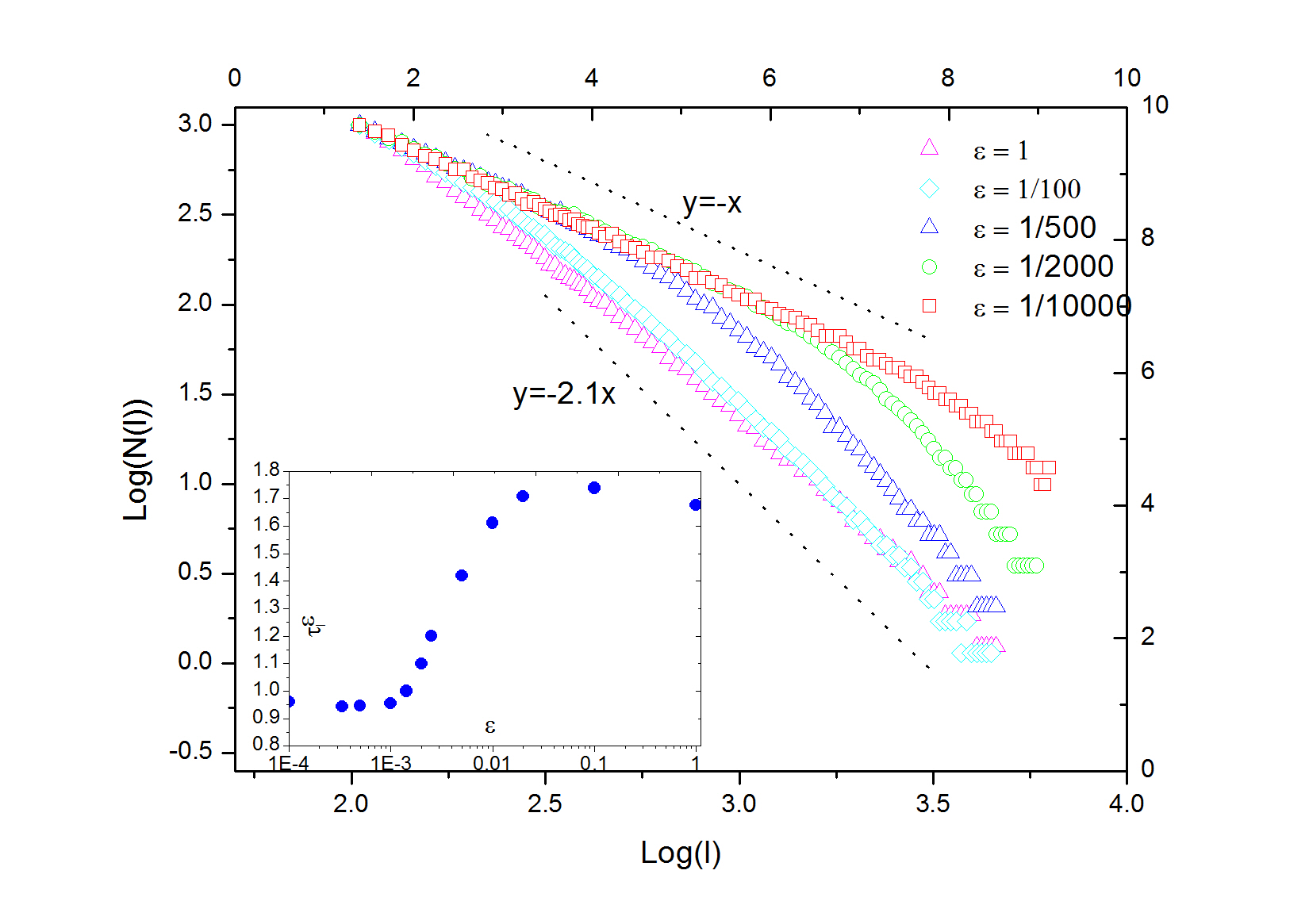}}
\caption{Log-Log plot of loop length distributions $N(l)$ versus $l$ of waves. The inner graph shows scaling of their slope in terms of mass.}
\label{loops}
\end{figure}
\begin{figure}
\centerline{\includegraphics[scale=.28]{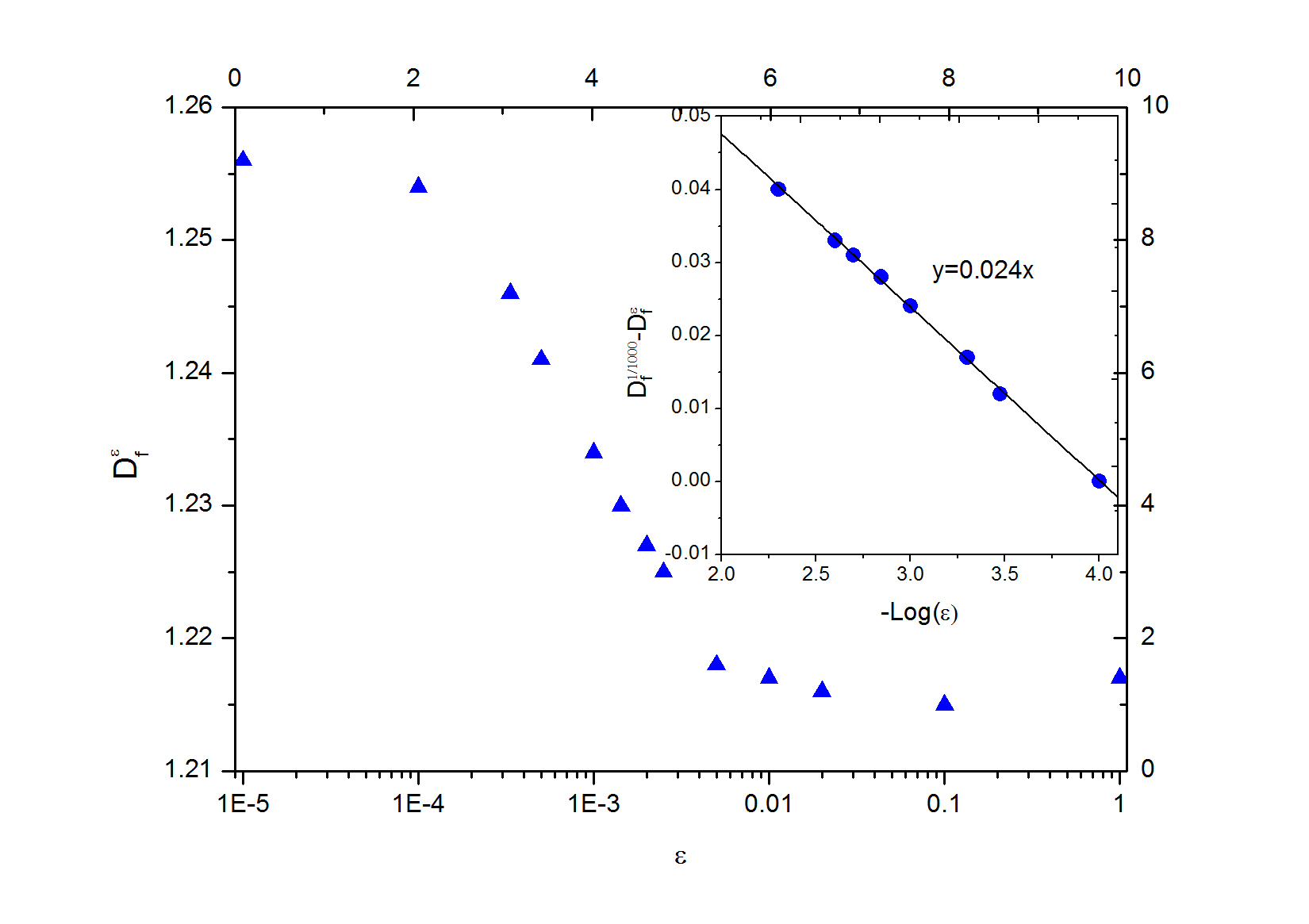}}
\caption{Fractal dimension of wave loops. The inner graph shows scaling of this in terms of mass.}
\label{fractal}
\end{figure}

In this section we numerically study the statistics of waves and avalanches of ASM to test its dependence on  $\epsilon$. For this, we have simulated over 100000 independent samples and obtained the domain walls between the toppled and untoppled sites on the honeycomb lattice of size 2048 $\times$ 2048. For the simulation, we considered the square lattice. The lattice and toppling rule hase been shown in FIG[\ref{sample}]. Consider the wave frontiers of toppled sites of ASM. We first study the statistics of the gyration radius of loops. In Fig[\ref{gyration}] the Log-Log plot of the distribution of gyration radius $N(r)$ versus gyration radius $r$ is sketched. For the BTW model ($\epsilon=0$), up to a length named as $r_{cut}^{(2)}$, it is seen that $N(r)\sim r^{-\tau_{r}}$ where $\tau_{r}\simeq 1\pm0.05$. For the lengths higher than $r_{cut}^{(2)}$ this distribution function falls off rapidly. $r_{cut}^{(2)}$ may be interpreted as the correlation length of the model and in the critical case, it is of order of the lattice size. By increasing $\epsilon$, another length scale is observed at which the behavior is changed. We name this length $r_{cut}^{(1)}$. As is shown in this figure, by increasing $\epsilon$, this point is changed and pushed along the origin. This point contains some interesting informations. Going from small lengths to large scales, we see two important lengths. In the small scales the curve is locally like BTW's up to the point $r_{cut}^{(1)}$. This suggests that the ultra violet (UV) properties of the perturbed model is given by the BTW model. In the vicinity of $r_{cut}^{(1)}$, the infra red (large scale) limit is reached and the behavior of the graph (more exactly, its slope) is smoothly changed to the new one. This and other figures (to be shown later) show that in the new regime the properties of the model is best fitted to the Manna model ($\epsilon=1$ in the figure). So it seems that the infra red (IR) limit of these curves is given by the Manna model and a cross-over takes place in between. The slopes of the graphs in the IR region is slightly different from the Manna model due to the finite size effects. In fact by enlarging the size of the lattice, we saw that the slopes in this region get closer to the Manna model (it is not shown here). The inner graph of this figure shows $\tau_{r}^{\epsilon}$ i.e. the slope of the graph for the small scales with respect to $\epsilon$. The horizontal axis is in logarithmic scale. As is seen, a smooth change of behavior takes place when we go from BTW to the Manna model. In the vicinity of the $\epsilon=0$ and $\epsilon=1$ the dominant behavior is the BTW's and the Manna's respectively and a jump takes place in between showing the mentioned cross-over. The same feature is seen in FIG[\ref{loops}] in which the distribution function of loop length '$N(l)$' is shown versus loop length '$l$'. In this case, as the $\epsilon$ gets non zero values, the deviation from the power law (governed on the first part of the graph) takes place at a characteristic length depending on $\epsilon$ value and this length decreases as $\epsilon$ increases. For $\epsilon=1$ the unique power law behavior is retrieved $N(l)\sim l^{-\tau_{l}^{\epsilon=1}}$ with $\tau_{l}^{\epsilon=1}=1.8\mp 0.1$. For $0<\epsilon<1$, the IR and UV properties the curves look like $\epsilon=1$ and $\epsilon=0$ cases respectively. A smooth transition for $\tau_{l}^{\epsilon}$ from BTW ($\tau_{l}^{0}=1$) to Manna ($\tau_{l}^{1}=1.8$) is seen in the inner graph of this figure. A similar calculations was done for the distributions of loop masses and the number of waves in the avalanches and loop sizes. We observed the same results as above.\\
By enlarging the size of the lattice, we saw that this graph was totally shifted to the left showing that enlarging the lattice size infinitely, the dominant behavior would be the Manna's. This tells us that this perturbation is relevant for the BTW fixed point and irrelevant for the Manna model.
For the scales much smaller than the correlation length, one expects that the system have a well defined fractal dimension. So we computed the fractal dimension of waves defined as $l\sim r^{D_{f}}$ and found that for finite $\epsilon$, there is a slight deviation from the critical fractal dimension (Note that surely for the lengths comparable with or larger than the correlation length, fractal dimension does not make sense, but to see the behavior of models with $\epsilon$'s near the Manna model, we calculated the fractal dimension for all $\epsilon$'s). The interesting feature, as is seen in FIG[\ref{fractal}], is that this quantity does scale with the logarithm of $\epsilon$  in the cross-over region i.e. $D_{f}^{0.0001}-D_{f}^{\epsilon}\sim$ Log($\epsilon$) and for large $\epsilon$'s it does not change and is nearly constant near the Manna model. This shows that this perturbation is irrelevant for the Manna fixed point. Table[I] shows some informations about the various exponents of the waves for various $\epsilon$. 

\begin{table}
\begin{center}
\begin{tabular}{|c|c|c|c|c|}
\hline $\epsilon$ & $r_{cut}^{(1)}$ & $\tau_{N_{w}}(\pm{0.02})$ & $\tau_{r}(\pm{0.02})$ & $D_{f}^{\epsilon}(\pm{0.005})$ \\ 
\hline $1$ & $---$ & $2.31$ & $1.93$ & $1.22$ \\ 
\hline $\frac{1}{500}$ & $29\pm{2}$ & $2.42$ & $0.95$ &  $1.23$ \\ 
\hline $\frac{1}{2000}$ & $59\pm{3}$ & $2.45$ & $0.94$ & $1.24$ \\ 
\hline $\frac{1}{10000}$ & $119\pm{4}$ & $2.46$ & $0.93$ & $1.25$ \\ 
\hline
\end{tabular}
\caption{The $r_{cut}^{(1)}$ and the exponents of gyration radius of waves $N(r)\sim{r^{\tau_{r}}}$, number of waves in an avalanche $N(N_{w})\sim N_{w}^{\tau_{N_{w}}} $, the fractal dimension of waves $\langle Log(l)\rangle = D_{f}^{\epsilon}\langle Log(r)\rangle$  in terms of $\epsilon$ for $L=2048$}.
\end{center}
\label{table1}
\end{table}

\section{DETERMINATION OF THE PERTURBING FIELD}  
\begin{figure}
\centerline{\includegraphics[scale=.40]{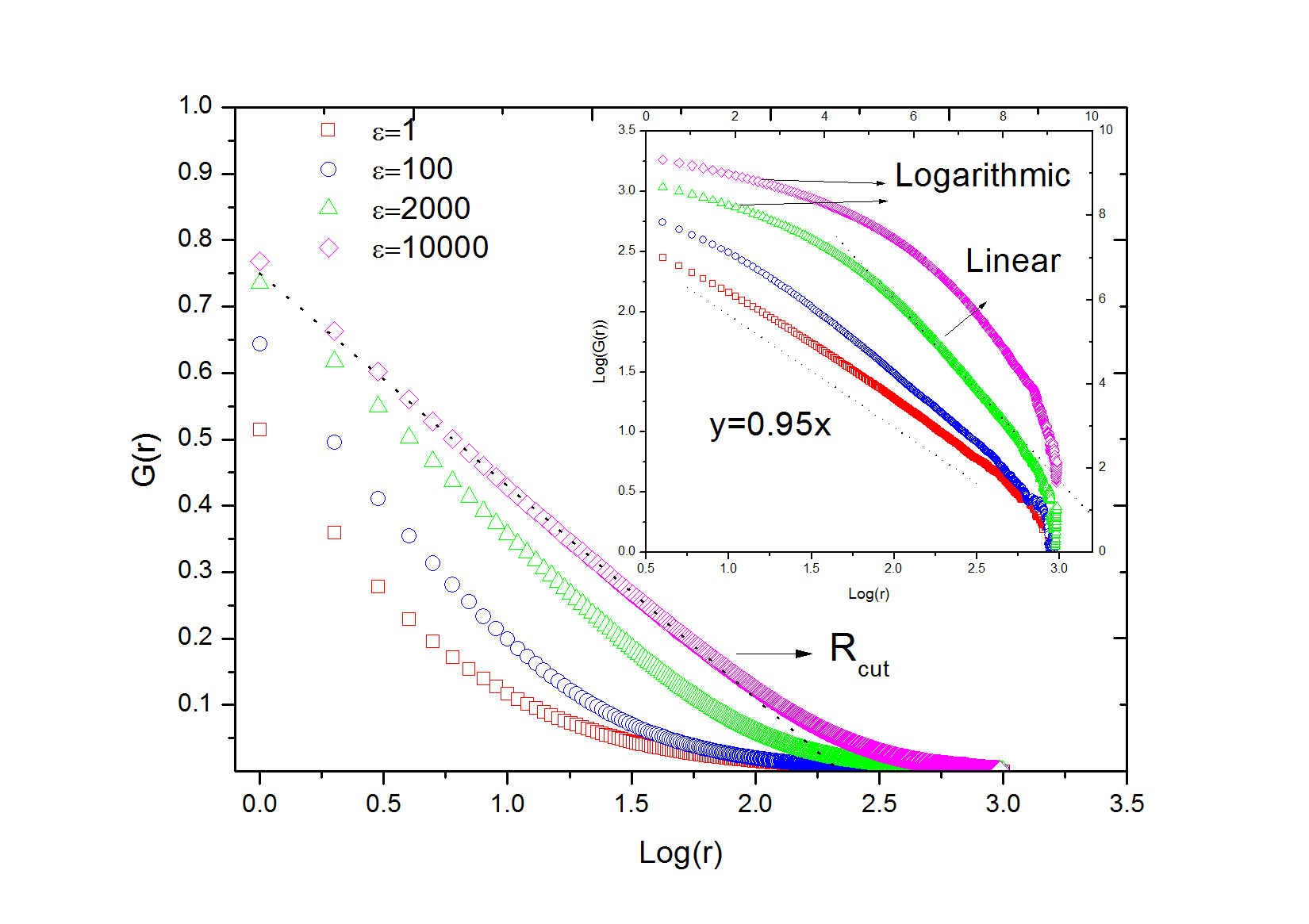}}
\caption{Green function vs Log(r). Inner graph shows the dependence of the slope of the first part of curves to the dis-
sipation. Note that some curves are added by a constant to have all the same first point.}
\label{Green}
\end{figure}

To determine the weight of the perturbing field we use two methods. First we directly try to determine this weight by using finite size effects. The second method we use is the determination of the Green function which depends directly to $\epsilon$.
\\
\newline \textbf{Green function method}; In this section using Green function of the perturbed model we present a method to compute the conformal weight of the perturbing field. As proved by Dhar \cite{Dhar3} the Green function of ASM is defined as follows: suppose that site $i$ is toppled by adding a grain. The Green function $G(\vert i-j\vert)$ is the number of topplings that occur in site $j$ (up to a normalization factor) and is proved that the form of this function in $2D$ is $G(r)\sim$ Log($r$) which Mathematically is the answer of the equation $(\frac{1}{r}\partial_{r}[r\partial_{r}])G(r,\acute{r})=\delta(r-\acute{r})$. When the action of a critical model is perturbed, the action of the model is modified. This modification may be of the form: 
\begin{equation}
S = S^{*} + \sum_{k=1}^{n}\lambda_{k}\int d^{2}x \varphi_{k}(x)
\label{action}
\end{equation}
in which $S^{*}$ is the action of the conformal field theory corresponding to the critical model (in this case the Eq[\ref{ghost}]) and $S$ is the perturbed action and $\lambda_{k}$ is the coupling constants of the perturbations. Then one expects that the differential equation of resulting Green function is also modified. In our case we can first search to find which equation the Green function does satisfy. Then we can observe that what is the conformal weight of the perturbing field. In Fig[\ref{Green}] the results of the simulations are sketched. In this figure the horizontal axis is in logarithmic scale. We see that for low $\epsilon$'s the resulting graph is logarithmic (up to a characteristic point $R_{cut}$ which is the finite size effect) as expected. But when $\epsilon$ increases, the graphs do not show this behavior. In the inner graph we have shown the Log-Log graphs and see that for $\epsilon=1$ the graph is power law $G(r)\sim r^{-x_{l}}$ with $x_{l}\simeq 1$. There is another way i.e. scaling argument, to study the Green function in IR regime. It is known that for a sandpile model the relation between the exponent of Green function and the exponent of the distribution function of wave gyration radius is \cite{KitLubGrasPri}:
\begin{equation}
-\tau_{r}+1+d_{f}^{(2)}-D=-x_{l}
\label{Green-scaling}
\end{equation}
where $D$ is the dimension of the space (here is $2$) and $\langle s\rangle \sim r^{d_{f}^{(2)}}$ where $s$ is the area of the loop and $r$ is its gyration radius. Here $d_{f}^{(2)}=2$. In the case $\epsilon=0$ ($\tau_{r}\simeq 1$) it is obvious that $x_{l}=0$ and the answer will be logarithmic. For the case $\epsilon=1$ also we have (we have calculated above that $\tau_{r}^{\epsilon=1}\simeq 2$) $x_{l}=1$ which confirms the calculation and the simulation.\\
 In the intermediate values of $\epsilon$ we see two different behaviors reflecting their IR and UV properties. For the lengths much smaller than the correlation length we see that they show similar features like the BTW model as expected. For the large scales (lengths much larger than the correlation length) they behave like the Manna model. This confirms our claim that the IR properties of the perturbed models best fit with the Manna model. Now we will search for the best differential equation which yield the mentioned properties. For large scales, the answers of this equation should be like the Manna model's i.e. $G(r)\sim r^{-1\pm 0.1}$ and for the small lengths, should have logarithmic form. We have observed that the best fit can be achieved by the following differential equation:
\begin{equation}
\frac{1}{r}\partial_{r}[r\partial_{r}G(r)]-g(\epsilon)\frac{1}{r}\partial_{r}[rG(r)]=\delta(r)
\label{Green-epsi}
\end{equation}
In this equation, $g(\epsilon)$ is some function of $\epsilon$. The numerical values of this function has been presented in Table[II]. For small values of $\epsilon$, we have $g(\epsilon)\sim \epsilon^{0.87\pm 0.04}$. We will use this form in the next subsection.

\begin{figure}
\centerline{\includegraphics[scale=.35]{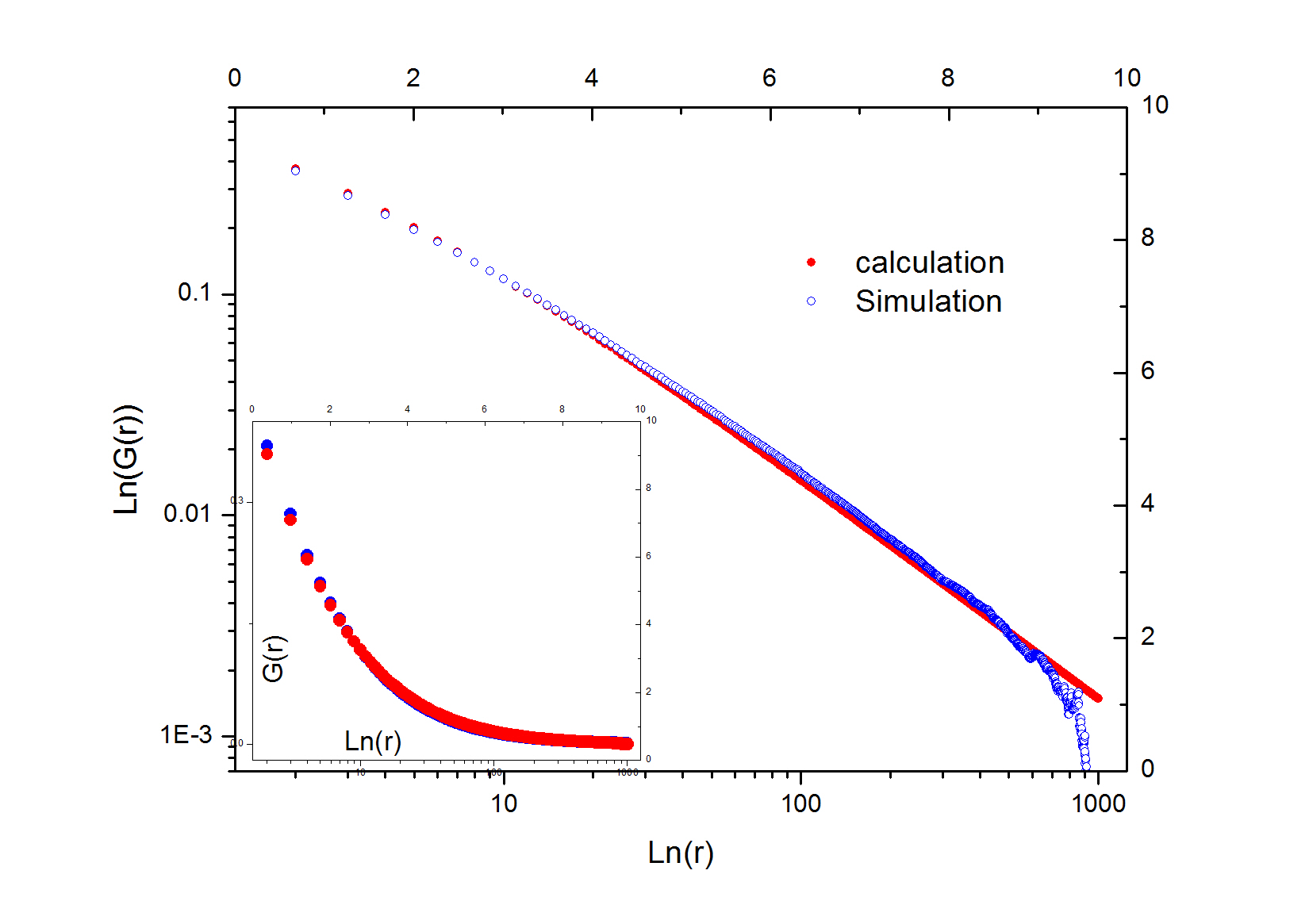}}
\caption{Green function vs Log(r). Inner graph shows the dependence of the slope of the first part of curves to the dis-
sipation. Note that some curves are added by a constant to have all the same first point.}
\label{Green2}
\end{figure}

\begin{table}
\begin{center}
\begin{tabular}{|c|c|}
\hline $\epsilon$ & $g(\epsilon)$ \\ 
\hline $1$ & $0.3\pm 0.02$ \\ 
\hline $0.1$ & $0.25\pm 0.02$ \\
\hline $0.02$ & $0.0.22\pm 0.02$ \\ 
\hline $0.0025$ & $0.11\pm 0.02$ \\ 
\hline $1/700$ & $0.08\pm 0.02$ \\ 
\hline $0.0005$ & $0.03\pm 0.02$ \\ 
\hline $0.0001$ & $0.007\pm 0.002$ \\  
\hline
\end{tabular}
\caption{The explicit amounts of $g(\epsilon)$ for different $\epsilon$'s} 
\end{center}
\end{table} 
The Eq.[\ref{Green-epsi}] properly yields the IR and UV properties of the Green function and for the intermediate lengths also fit with the simulation. For example in FIG[\ref{Green2}] we have shown the simulation and calculated (from Eq[\ref{Green-epsi}] with $g(\epsilon=0.1)=0.25$) results which have been fitted well. From the extra term $\frac{1}{r}\partial_{r}[rG(r)]$ we are led to the very important conclusion that the conformal weight of the perturbing field is $2\Delta =1$. This statement will be further tasted in the next subsection in which we try to yield $\Delta$ from RG arguments.\\

\begin{figure}
\centerline{\includegraphics[scale=.35]{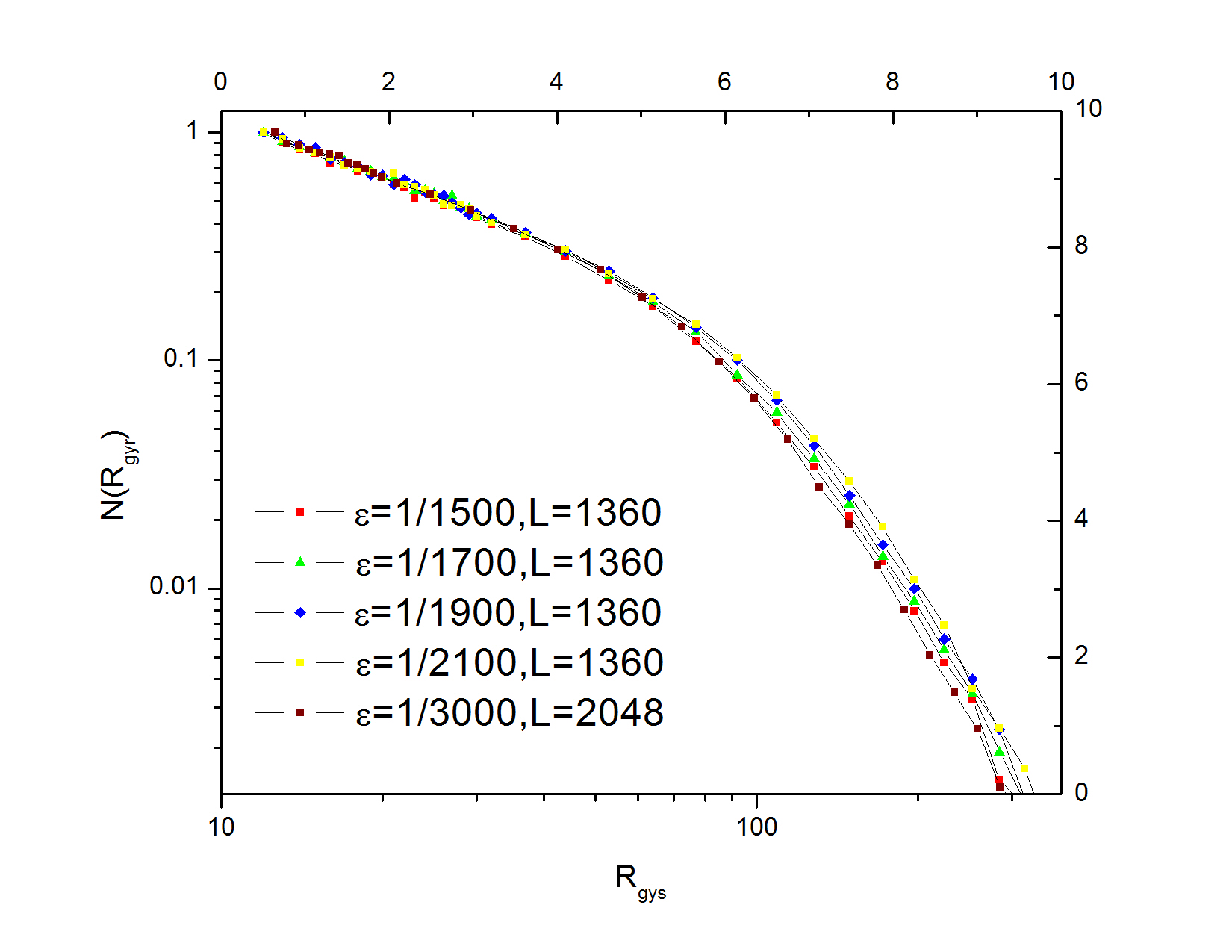}}
\caption{Determination of the weight of perturbing field. The curve is best fitted to $\epsilon=1/1500$}
\label{delta}
\end{figure}

\textbf{RG equation}; In the previous subsection we obtained the scaling of $g(\epsilon)$ versus $\epsilon$. In this subsection, using this construction and the RG arguments, we check the validity of this claim. We use the equation that govern the off-critical conformal field theory (in vicinity of the fixed point). Suppose that the action of the theory is:
It can be easily proved that the RG equation for this couplings is \cite{Mussardo}:
\begin{equation}
\frac{\partial g_{k}}{\partial b}=2(1-\Delta_{k})g_{k}-\pi \sum_{i,j}C_{i,j,k}g_{i}g_{j}
\label{RG}
\end{equation}
where $\lambda_{k}=g_{k}a^{2(1-\Delta_{k})}$, $\Delta_{k}$ is the conformal weight of the perturbing field $\varphi_{k}$, $b$ is the length scale, $a$ is the lattice constant and $C_{i,j,k}$ is the fusion coefficient of the fields $\varphi_{i}$ and $\varphi_{j}$ i.e. $\varphi_{i} \varphi_{i} =\sum_{k} C_{i,j,k} \varphi_{k}$. In our case, we have single coupling constant $g(\epsilon)$ and we can let $g(\epsilon)^{2}\simeq 0$. So we have($g(\epsilon)\sim \epsilon^{n}$ with $n=0.87\pm 0.04$ for small $\epsilon$'s):
\begin{equation}
\frac{\delta g(\epsilon)/g(\epsilon)}{\delta b}=n\frac{\delta \epsilon/\epsilon}{\delta b}\simeq 2(1-\Delta)
\label{RG-epsi}
\end{equation}
where $\Delta$ is the weight of the perturbing field. For using the RG equation [RG], we can set $a\rightarrow a(1+\delta b)$ or equivalently let the size of lattice $L\rightarrow \frac{L}{1+\delta b}\equiv \acute{L}$. So the correlation length of the system $\xi\rightarrow \frac{\xi}{1+\delta b}$. To use Eq.[\ref{RG-epsi}], it is sufficient to repeat the simulation for $\acute{L}$ and see for which $\epsilon$ two cases do match. In Fig[\ref{delta}] we have presented the result for $L=2048$ and $\delta b=0.5$ and $\epsilon_{2}=1/3000$. We see that the best fit is done for $\epsilon_{1}=1/1500$. Then we have

\begin{equation}
2\Delta=2-n\frac{\delta \epsilon/\bar{\epsilon}}{\delta b}=2-\frac{0.87}{0.53}0.65\simeq 0.94
\label{solve}
\end{equation}

Where $\bar{\epsilon}$ is the averaged value of the $\epsilon$ in this interval. We have done this test on many such samples with various lengths and $\epsilon$'s and found the best values for $\Delta$ are $0.9\leq 2\Delta\leq 1.05$ in agreement with the obtained result in the previous subsection.\\
\section{SCHRAMM-LOEWNER EVOLUTION}\label{SLE}
Critical behaviour of the two dimensional statistical models can be described by their geometrical features. In fact instead of studying the local observables, we can focus on the interfaces of two dimensional models. These domainwalls are some non-intersecting curves which directly reflect the status of the system in question and supposed to have two properties: conformal invariance and the domain Markov property\cite{Cardy2}. Schramm- Loewner Evolution is the candidate to analyze these random curves by classifying them to the one-parameter classes (SLE$_{\kappa}$). 
Let us denote the upper half plane by  $H$ and  $\gamma_{t}$ as the SLE trace i.e.  $\gamma_{t}=\lbrace z\in H:\tau_{z}\leq t \rbrace$ and the hull $K_{t}=\overline{\lbrace z\in H:\tau_{z}\leq t \rbrace}$.   $SLE_{\kappa}$ is a growth processes defined via conformal maps which are solutions of Loewner's equation:
\begin{equation}
\partial_{t}g_{t}(z)=\frac{2}{g_{t}(z)-\xi_{t}}
\end{equation}
Where the initial condition is $g_{t}(z)=z$  and $\xi_{t}=\sqrt{\kappa}B_{t}$ is a real valued smooth function. For fixed $z$, $g_{t}(z)$ is well-defined up to time $\tau_{z}$ for which $g_{t}(z)=\xi_{t}$. The complement $H_{t}:=H\backslash{K_{t}}$ is simply connected and $g_{t}(z)$ is the unique conformal mapping $H_{t}\rightarrow{H}$ with $g_{t}(z)=z+\frac{2t}{z}+O(\frac{1}{z^{2}})$ as $z\rightarrow{\infty}$ that is known as hydrodynamical normalization. One can retrieve the SLE trace by $\gamma_{t}=\lim_{\epsilon\downarrow{0}}g_{t}^{-1}(\xi_{t}+i\epsilon)$. There are phases for these curves, $2\leq\kappa\leq{4}$ the trace is non-self-intersecting and it does not hit the real axis; $k_{t}=\gamma_{t}$. This is called "dilute phase". But for $4\leq\kappa\leq{8}$, the trace touches it self and the real axis so that a typical point is surely swallowed as $t\rightarrow\infty$ and $K_{t}\neq\gamma_{t}$. This phase is called "dense phase". However, there is an important property: The frontier of $K_{t}$ i.e. the boundary of $H_{t}$ minus any portions of the real axis is a simple curve and is locally SLE$_{\tilde{\kappa}}$ with $\tilde{\kappa}=\frac{16}{\kappa}$. This duality links models in dilute phase to one model in the dense phase and vice versa e.g. the ASM ($\kappa=2$) to the Uniform Spanning Tree (UST) ($\kappa=8$).
The main question "what is the relation between SLE and CFT" is answered by M. Bauer and D. Bernard\cite{BauBer}. They showed that the boundary condition changing (bcc) operator in SLE correspond to a degenerate field with a vanishing descendant at level two and conformal weight $h_{1;2}=\frac{6-\kappa}{2\kappa}$ in CFT with central charge $c_{\kappa}=\frac{(6-\kappa)(3\kappa-8)}{2\kappa}$.
\newline 
\newline \textbf{SLE Out of criticality}; Now consider the system out of criticality. In this case the conformal invariance of the system is broken and the the system correlation length $\zeta$ will have a crucial role in statistical properties of the random curves. So if we apply the Loewner uniformizing map, the resulting domains are not more equivalent due to absence of conformal invariance. At scales much smaller than the correlation length, i.e. in the ultraviolet
regime, the deviation from criticality is small, and the interface should look locally like the critical interface. This means that over short time periods, the off-critical driving function $\xi_{t}^{\zeta}$ should not be much different from its critical counterpart. In the other hand, at large scales (with respect to $\zeta$), i.e. in the infrared regime, the interface may look like another SLE with a new $\kappa_{ir}$. The reason is; when we integrate out the small distances to reach the large distance properties, the regions which is formed by SLE trace for lengths smaller than the correlation length (with diffusivity $\kappa_{UV}$), may be seen as points that the SLE trace with the new diffusivity ($\kappa_{ir}$) corresponding to the new infra red fixed point crosses. The example is Ising model. At criticality $\kappa=3$, but if the temperature is raised above the critical point, renormalization group arguments indicate that at large scale the interface looks like the interface at infinite temperature i.e. percolation with $\kappa_{ir}=6$\cite{BauBerKal}. 
\newline Now consider a curve that starts from origin and end on a point on real axis ($x_{\infty}$). Then by using the map $\phi={x_{\infty}z}/{(x_{\infty}-z)}$, one can send the end point of the curve to the infinity. In this respect, the function $h_{t}=\phi{o}g_{t}{o}\phi^{-1}$ describes chordal SLE. It is easy to show that the equation governing on $h_{t}$ is:
\begin{equation}
\partial_{t}g_{t}=2/(\lbrace{\acute{\phi}(g_{t})(\phi(g_{t})-\xi_{t})}\rbrace).
\end{equation}
This mapping is not hydrodynamically normalized i.e. it does not fix the infinity, instead it fixes the ending point of the curve.
\section{numerical results; schramm-loewner evolution}
\begin{figure}
\centerline{\includegraphics[scale=.30]{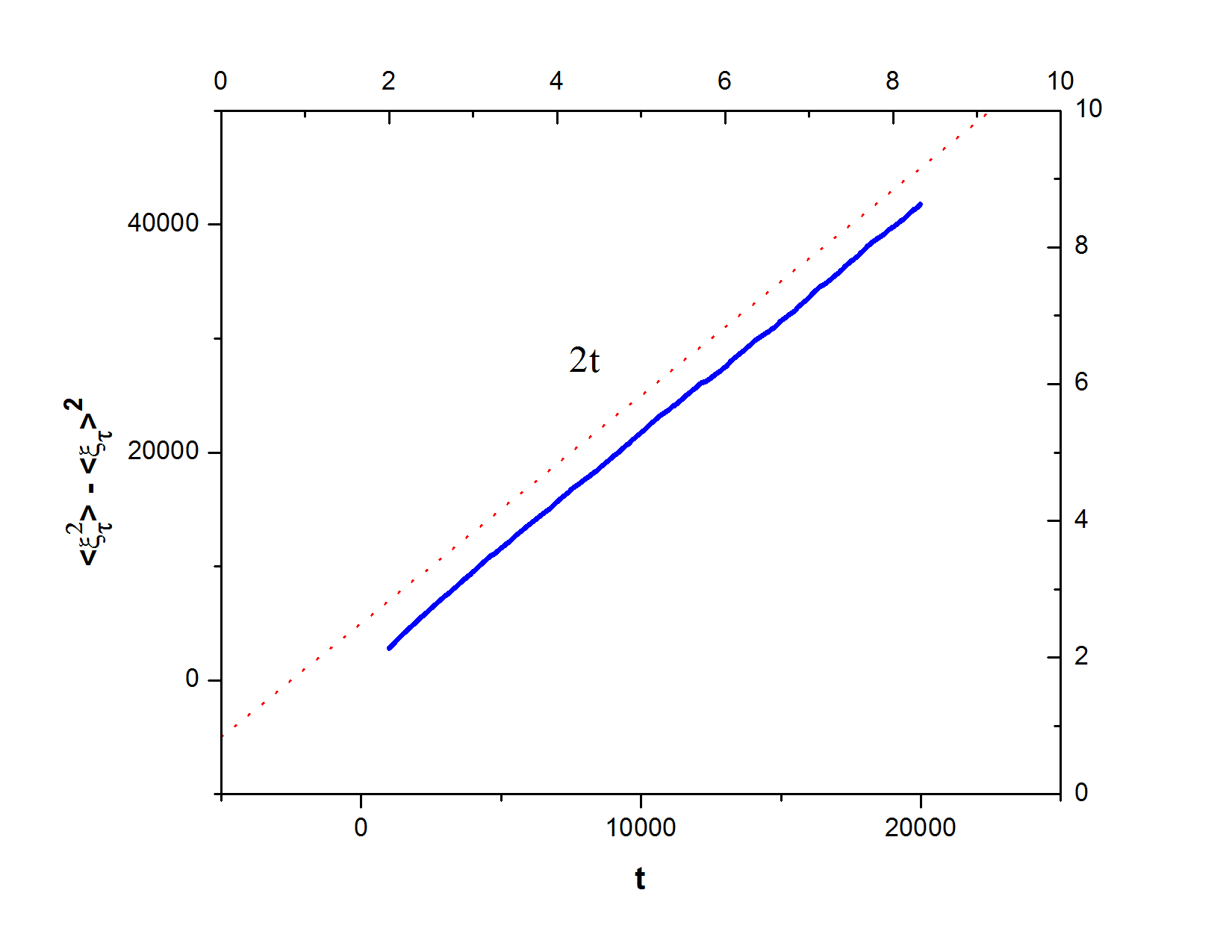}}
\caption{The averaged squared $\xi_{t}$ versus $t$ shows the diffusivity $\kappa=2.0\pm{0.1}$.}
\label{kappa2}
\end{figure}

\begin{figure}
\centerline{\includegraphics[scale=.35]{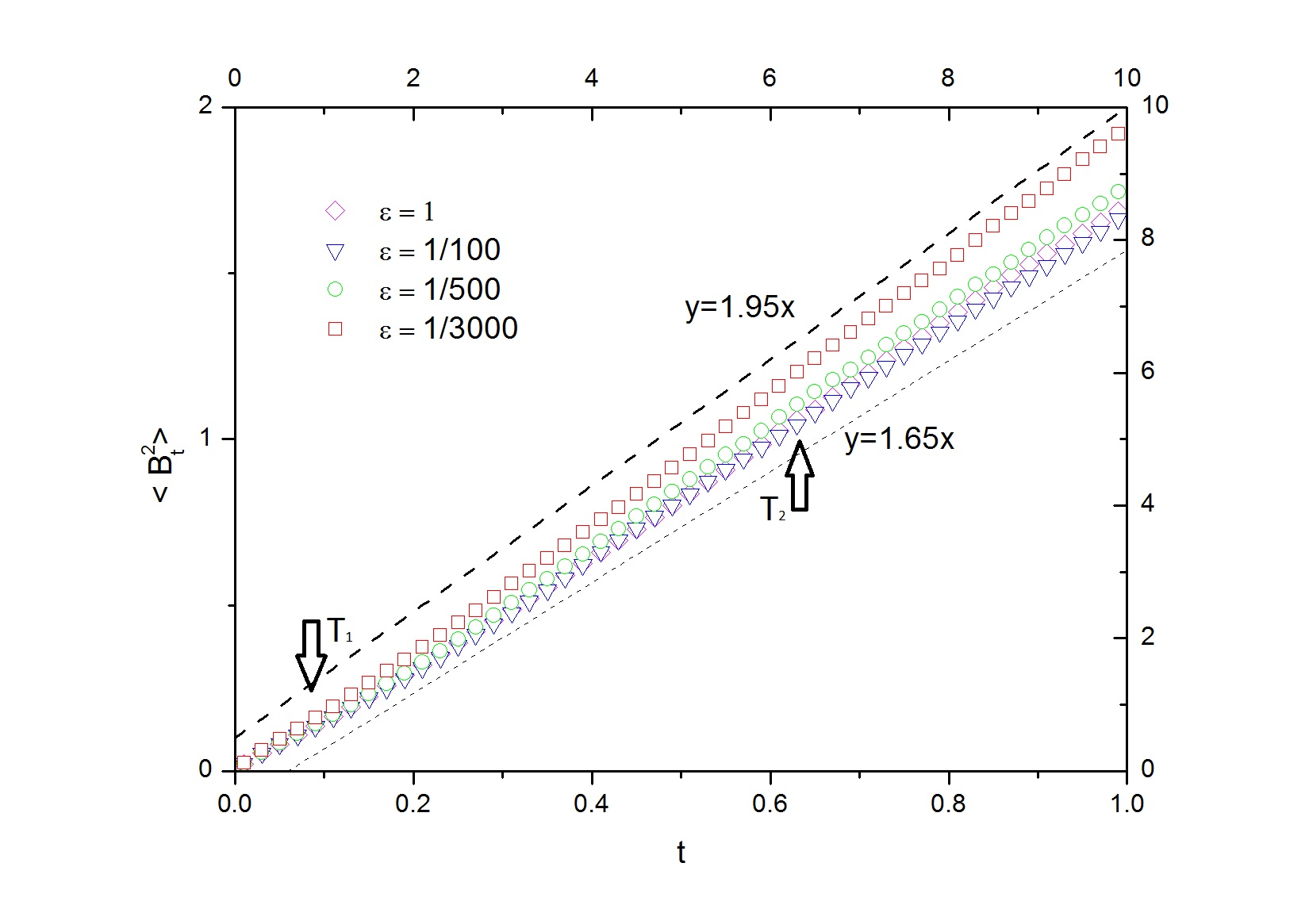}}
\caption{$(\langle\xi_{t}^{2}\rangle-\langle\xi_{t}\rangle^{2})$ versus $t$ for different dissipations.}
\label{total}
\end{figure}

\begin{figure}
\centerline{\includegraphics[scale=.30]{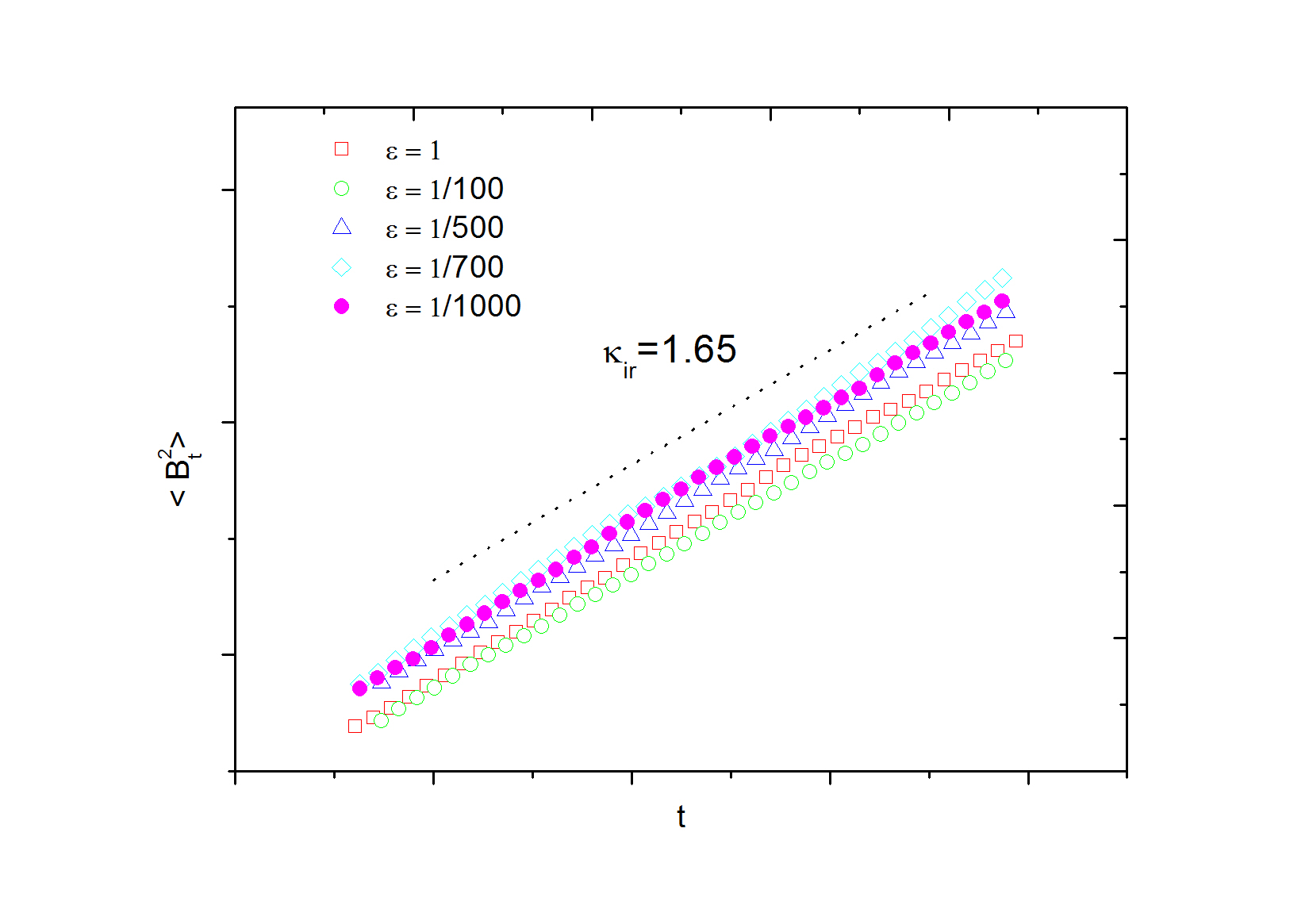}}
\caption{$\kappa_{ir}$ for various dissipations in region II.}
\label{k_ir}
\end{figure}

In this section we present some numerical results obtained by applying SLE on the critical and off-critical abelian sindpile model. The the frontier of avalanches form the set of loops with discrete points. As in the chordal SLE, the curve goes from a point of the real axis to a point in the infinity and here we have loops, we are to use a trick to generate desired curves. Having these loops, one can cut them with a straight line to generate interface curves starting from the origin and ending at some point on the real axis ($x_{\infty}$). Then using the map $\phi={x_{\infty}z}/{(x_{\infty}-z)}$, that fixes the origin and sends the ending point ($x_{\infty}$) to the infinity, we will have a curve on the upper half plane. Then by applying the chordal SLE formalism and the proper uniformizing map step by step, one can emerge $\xi_{t}$ for these discrete curves. The essential assumption is that $\xi_{t}$ is partially (in each interval) constant, then it can be easily proved that the mapping that can be used to uniformize the curves is\cite{BBCF}:

\begin{equation}
G_{t}(z)=x_{\infty}\frac{\eta{x_{\infty}}(x_{\infty}-z)+a(\xi_{t},z)}{x_{\infty}^{2}(x_{\infty}-z)+a(\xi_{t},z)}
\end{equation}

where $\eta=\phi^{-1}(\xi)$ and $a(\xi_{t},z)=\sqrt{x_{\infty}^{4}(z-\eta)^{2}+4t(x_{\infty}-z)^{2}(x_{\infty}-\eta)}$. This mapping uniformizes a semicircle that is extended from $\eta$ to $x_{\infty}$ and by demanding that this semicircle involve $z_{1}$ one obtain: 

\begin{equation}
\eta_{t}=\dfrac{Re(z_{1})x_{\infty}-[Re(z_{1})]^{2}-[Im(z_{1})]^{2}}{x_{\infty}-Re(z_{1})}
\end{equation}
\begin{equation}
t=\dfrac{Re(z_{1})^{2}x_{\infty}^{4}}{4(Re(z_{1})-x_{\infty})^{2}+Im(z_{1})]^{2}}
\end{equation}
FIG[\ref{kappa2}] contains the graph $\langle\xi_{t}^{2}\rangle-\langle\xi_{t}\rangle^{2}$ versus $t$ for the critical case. As it is explicit in the graph, $\xi_{t}$ has the expected behaviour: $\langle\xi_{t}\rangle\simeq{0}$ and $(\langle\xi_{t}^{2}\rangle-\langle\xi_{t}\rangle^{2})=\kappa{t}$ with $\kappa=2.0\pm{0.1}$. We note that the initial portion ($0<t<1000$) of the graph is different from the remaining and has been ignored. The reason is that for these times, the effect of finite size (lattice constant) on the curve growth is important as the size of the curve is comparable with it. So in the remaining of the paper we will ignore this portion. We however are not concerned about the effect of the size of the system, because it take very long time for such a fractal to have a linear size of the system order.\\
For the off-critical model, as stated in Sec. \ref{SLE}, we have two important scale limits. For small scales (scales much smaller than $r_{cut}^{(1)}$) the interface should look locally like the critical interface at UV fixed point. At large scales however, the interface may behave like a SLE corresponding to the IR fixed point with $\kappa_{ir}$ . We consider the perturbed ASM as described bove and analyze the resulting driving function. The important quantity which can be extracted from driving function is $\kappa$ which in the critical case is obtained from the relation $\langle\xi_{t}^{2}\rangle-\langle\xi_{t}\rangle^{2}=\kappa t$. In the off-critical case we may observe two slopes for the graph: one for UV region (the resulting $\kappa$ should not be much different from the critical one) and another for IR region. In between the curve may have complex behaviors. In FIG[\ref{total}] $\langle\xi_{t}^{2}\rangle-\langle\xi_{t}\rangle^{2}$ versus $t$ is shown for some $\epsilon$'s. The slope of the graph for the critical ($\epsilon=0$) one is $\kappa = 1.95 \pm 0.1$ in agreement with other numerical results\cite{Saberi}. When  $\epsilon$ becomes non-zero, the graphs does not show simple linear behavior. As is seen from this figure, there are two transition points: The first one is the earliest time in which the graph separates from the critical one i. e. is the first transition point from UV to the "cross over" region (in the FIG[\ref{total}] is shown as $T_{1}$). The next ($T_{2}$) is the transition from the "cross over region" to the the IR region. In this region, the slope is the same as the Manna's i. e. $\epsilon=1$.These transition points depend on $\epsilon$ and increase as $\epsilon$ decreases and in the case $\epsilon=0$ become infinite. We can investigate the behaviour of the curves at UV and IR regions well separated from the crossover region. The result is that the curves are linear in each region with the same slope in each region. These feature for IR regime has been shown in Fig[\ref{total}] and magnified in FIG[\ref{k_ir}] in which is seen that all have the same slope $\kappa_{ir}=1.65\pm 0.1$.\\


\section{Conclusion}
     In this paper, we analyzed the statistics of wave and avalanche frontiers of continuous random anisotropic ASM. The BTW model corresponds to the perturbation parameter $\epsilon=0$ and Manna model to $\epsilon=1$. It has been shown that a cross over takes place between these two models. We studied the behavior of some statistical observables and found the conformal weight of the perturbing field by two methods: Green function and RG arguments. Each of them confirm that the weight $x=1$. Using SLE for the geometric curves of the perturbed model we showed that there are two important length scales in which the corresponding SLE parameter '$\kappa$' is different. These scales are determined with respect to the correlation length. Using SLE, we found numerically that for the scales much smaller than the correlation length, the curves have the same properties as the UV critical model (BTW) with nearly the same $\kappa$. For scales much larger than it, also we found that the curves acquire the new $k_{ir}\simeq 1.65$.

\end{document}